\documentclass[a4paper]{article}
\usepackage{Odyssey2018}
\usepackage{epsfig,amssymb,amsmath}
\ninept

\usepackage[utf8]{inputenc}
\usepackage{amsmath,graphicx}
\usepackage{amssymb,amsmath,epsfig,url,mathrsfs} \usepackage{algorithm,algorithmic}
\usepackage[usenames, dvipsnames]{color}
\usepackage[framemethod=TikZ]{mdframed}
\usepackage{xcolor}
\usepackage{colortbl}
\usepackage[colorinlistoftodos]{todonotes}
\usepackage{hyperref}
\usepackage{wasysym}

\usepackage{graphicx}
\usepackage{subfigure}

\renewcommand{\vec}[1]{\boldsymbol{\mathrm{#1}}}
\newcommand{\mtx}[1]{\boldsymbol{\mathrm{#1}}}

\renewcommand{\vec}{\boldsymbol} 

\usepackage{amsthm}

\makeatletter
\newcommand{\mathleft}{\@fleqntrue\@mathmargin0pt}
\newcommand{\mathcenter}{\@fleqnfalse}
\makeatother
 
\theoremstyle{definition}

\hypersetup{
   unicode=false,          
   pdftoolbar=true,        
   pdfmenubar=true,        
   pdffitwindow=false,     
   pdfstartview={FitH},    
   pdftitle={My title},    
   pdfauthor={Author},     
   pdfsubject={Subject},   
   pdfcreator={Creator},   
   pdfproducer={Producer}, 
   pdfkeywords={keyword1} {key2} {key3}, 
   pdfnewwindow=true,      
   colorlinks=true,       
   linkcolor=blue,          
   citecolor=blue,        
   filecolor=magenta,      
   urlcolor=blue           
}

\mdfdefinestyle{MyFrame}{%
    linecolor=black,
    outerlinewidth=.1pt,
    roundcorner=1pt,
    innertopmargin=\baselineskip,
    innerbottommargin=\baselineskip,
    innerrightmargin=5pt,
    innerleftmargin=-10pt,
    backgroundcolor=gray!10!white}
\makeatletter
\def\blfootnote{\xdef\@thefnmark{}\@footnotetext}
\makeatother

\title{A Spoofing Benchmark for the 2018 Voice Conversion Challenge:\\Leveraging from Spoofing Countermeasures for Speech Artifact Assessment}

\makeatletter
\def\name#1{\gdef\@name{#1\\}}
\makeatother
\name{{\em Tomi Kinnunen$^1$, Jaime Lorenzo-Trueba$^2$, Junichi Yamagishi$^{2,3}$, Tomoki Toda$^4$,}\\
      {\em Daisuke Saito$^5$, Fernando Villavicencio$^6$, Zhenhua Ling$^7$}}

\address{$^1$ University of Eastern Finland, Joensuu, Finland\\
$^2$ National Institute of Informatics, Tokyo, Japan 
$^3$ University of Edinburgh, UK \\
$^4$ Nagoya University, Nagoya, Japan 
$^5$ University of Tokyo, Tokyo, Japan \\
$^6$ ObEN, Pasadena, USA 
$^7$ University of Science and Technology of China, Heifei, China \\ 
{\small \tt vcc2018@vc-challenge.org} }

\begin{document}
\maketitle
\blfootnote{\textcolor{magenta}{This is a \textbf{corrected} version of the \emph{Odyssey 2018} publication with the same title and author list. A bug (partial training-test overlap in \emph{bona fide} trials) was identified afterwards, leading to underestimated EERs on VCC'18 base and VCC'18 (challenge entries) data. The overlapped trials were removed and the affected Tables (1, 2, 3 and 4) and Figs. (1 and 2) and their descriptions in the text were updated. The overall trends and conclusions remain unchanged. Date of the document: \today.}}

\begin{abstract}
Voice conversion (VC) aims at conversion of speaker characteristic without altering content. Due to training data limitations and modeling imperfections, it is difficult to achieve believable speaker mimicry without introducing processing artifacts; performance assessment of VC, therefore, usually involves both speaker similarity and quality evaluation by a human panel. As a time-consuming, expensive, and non-reproducible process, it hinders rapid prototyping of new VC technology. We address artifact assessment using an alternative, objective approach leveraging from prior work on spoofing countermeasures (CMs) for automatic speaker verification. Therein, CMs are used for rejecting `fake' inputs such as replayed, synthetic or converted speech but their potential for automatic speech artifact assessment remains unknown. This study serves to fill that gap. As a supplement to subjective results for the 2018 Voice Conversion Challenge (VCC'18) data, we configure a standard constant-Q cepstral coefficient CM to quantify the extent of processing artifacts. Equal error rate (EER) of the CM, a confusability index of VC samples with real human speech, serves as our artifact measure. Two clusters of VCC'18 entries are identified: low-quality ones with detectable artifacts (low EERs), and higher quality ones with less artifacts. None of the VCC'18 systems, however, is perfect: all EERs are $<30\%$ (the `ideal' value would be 50\%). Our preliminary findings suggest potential of CMs outside of their original application, as a supplemental optimization and benchmarking tool to enhance VC technology. 
\end{abstract}

\section{Introduction}


\emph{Voice conversion} (VC) \cite{Mohammadi2017-vc-overview,stylianou09-VC} aims at conversion of speaker characteristic without altering the speech content. Typical use of VC technology includes applications in entertainment industry, such as customizing  artificial voices for audio-books and games. In such applications, the VC samples are optimized for human listeners. In the recent past, thanks to technological advances in both VC and automatic speaker verification (ASV) technology, VC finds also frequent use in assessing ASV system vulnerability against intentional circumvention (spoofing) \cite{Kinnunen2012-vulnerability}. In this case, the VC samples are prepared for the ASV system. In both cases, the goal is that the VC samples manage to make the primary observer (either a human or a machine) to believe they are observing a certain targeted speaker that is different from the source speaker. But human perception and machine perception are different, and in the case of ASV and its spoofing, machine perception is more relevant.

Even if the VC technology itself has evolved steadily over the years \cite{toda2007voice,SAITO20172017EDL8034,Kobayashi2017}, the evaluation methods of VC are more varied compared to tasks such as ASV or automatic speech recognition. The primary evaluation methods are perceptual tests since the target of the above applications is normally human perception. In addition, log-spectral distortion and cepstral distortion (between converted and target utterances) are also used as supplementary information. At this moment, we lack of universally adopted objective performance measures. Furthermore, there was no standard database until recently. 

Given the situation, \cite{Toda2016-VCC-challenge} launched a \emph{Voice Conversion Challenge} (VCC) series in 2016, with a follow-up in 2018\footnote{\url{http://www.vc-challenge.org/}, data available at \url{http://dx.doi.org/10.7488/ds/2337} since April 10, 2018.} organized by the authors of this study \cite{Lorenzo2018-VCC18}. The primary methodology of the evaluation of VC systems, including VCC, is a perceptual test of various VC systems trained on a common corpus. The perceptual test usually involves both speaker similarity and quality evaluation by a human panel since it is difficult to achieve convincing speaker transformation without introducing processing artifacts due to training data limitations and modeling imperfections. The VCC series provide results of large-scale perceptual tests that compare many different types of VC systems on a comparable basis. This is helpful towards understanding human perception and optimization strategies employed by listeners. On the other hand, listening test results do not directly reflect spoofing capability and we do not know how the results of the perceptual test are related to machine perceptions, that is, ASV and its spoofing. 

With the above motivations in mind, the present study accompanies \cite{Lorenzo2018-VCC18} that provides details of the 2018 challenge data, analysis of the submitted systems and the perceptual results. We provide supplemental objective quality results on the degrees of artifacts that each of the submitted systems have. Even if our experiments are framed to the context of the latest VCC'18 challenge, our contribution is that of a novel objective speech artifact assessment leveraging from the rapidly emerging topic of \emph{spoofing countermeasures}. In the context of ASV, \emph{spoofing} refers to intentional circumvention of the ASV system to obtain illegitimate access as another targeted user \cite{Wu2015-spoofingsurvey} --- VC technology being a representative example. ASV vulnerability due to spoofing 
has been known for about two decades \cite{Pellom1999-voice-altered} but has gained momentum only relatively recently with increased interest towards ASV deployment for user authentication, as well as availability of common evaluation resources \cite{Wu2017-asvspoof,Korshunov-BTAS2016overview} to enable meaningful comparisons of different spoofing countermeasures.

There has been continued research towards generalized spoofing countermeasures to detect spoofing attacks more accurately. As a result, several advanced front-end \cite{Sahid2015-features,Todisco2016-cqcc,Sriskandaraja-scattering-2017,Patel15-combining} and machine learning \cite{Lavrentyeva2017} oriented techniques have been developed for the task of detecting the presence of a spoofing attack in a given audio segment. The task is framed as a hypothesis testing problem with \emph{bona fide} (legitimate human speech) hypothesis as the null hypothesis and \emph{spoof} as an alternative hypothesis. The exact definition of the latter depends on the type of the spoofing attack (\emph{e.g.} VC or replay attack).

If the detector is carefully optimized and the probability distributions of bona fide and spoof classes are sufficiently distinct, one can detect the attacks. But if the spoofed samples resemble too closely real human speech, the detector can mistakenly classify them as bona fide speech. Therefore, the number of errors made by a spoofing countermeasure for a given batch of test files is associated with how closely the spoof samples resemble the bona fide samples. In specific, the spoofing countermeasure gauges the amount of speech artifacts that only the spoof samples have (regardless whether the artifacts are audible to a human or not) and tell us how close the spoof samples are compared to the bona fide samples. Therefore we hypothesize that this is useful for the automatic assessment of speech artifacts produced by VC process and may be used as one of objective performance measures. Therefore, this study compares the performance of the spoofing countermeasure with subjective quality evaluation results obtained in VCC'18 and investigate how they are related to each other.



\section{Subjective quality of converted voices}

Suppose we have a patch of $N$ source speaker utterances $\mtx{X} = \{\mathcal{X}_1,\dots,\mathcal{X}_N\}$ processed through $S$ voice conversion systems $s=1,\dots,S$. We denote the utterances converted by system $s$ by $\mtx{Y}^s = \{\mathcal{Y}^s_1,\dots,\mathcal{Y}^s_N\}$ and use $\mtx{Y} = \cup_{s=1}^S \mtx{Y}^s$ to denote all the $N \cdot S$ conversions. For the purpose of comparing the alternative VC systems in terms of speech quality, the samples $\mtx{Y}$ are collectively listened to by a cohort of human \emph{observers}, $\mathcal{O} = \{O_1,\dots,O_L\}$, each of who outputs an \emph{opinion score} for a subset of samples indexed by $\alpha_i$ for observer $i$. Note that not all the observers necessarily listen to the same utterances, nor necessarily listen to even the same number of samples. The process of obtaining the opinion score of observer $i$ for utterance $\mathcal{Y} \in \mtx{Y}$ can be thought of as evaluating an abstract, non-deterministic and possibly time-varying, function $h(\mathcal{Y},O_i)$, that models human listening mechanism of $O_i$. It depends on many factors such as the listener's life experience and concentration, the listening environment, audio equipment used, and familiarity with the language. We do not have access to the internals of $h(\cdot,O_i)$ but only its observed output, in this study the standard 5-point rating scale ranging from 1 (lowest quality) to 5 (highest quality). 

Because of random variation in the outputs, caused by differences in $h(\cdot,O_i)$, one represents the results of the listening panel in an averaged form. The well-known population summary measure, \emph{mean opinion score} (MOS), is computed for system $s$ by $\text{MOS}_s = (1/L_s) \sum_{n=1}^N \sum_{i=1}^L h(\mathcal{Y}^s_n,O_i)$, where $L_s$ is the total number of opinion scores obtained for the samples of system $s$, and where we assign a dummy value $h(\mathcal{Y},O_i)=0$ if listener $i$ did not rate $\mathcal{Y}$. The higher the MOS value, the higher quality the samples of system $s$. The definition of 'quality' is also subjective. No instruction on what high-quality or low-quality means is normally given. 

\section{Proposed objective artifact assessment using spoofing countermeasures}

Here we want to construct a model that automatically scores the amount of speech artifacts. The artifacts may be audible or non-audible\footnote{Hence the aim of the measure is not to approximate the subjective quality judgement.}. 

\subsection{Obtaining machine scores}

With the objective artifact estimation, the problem setup is the same as above: given $\mtx{Y}^s$, we want to obtain a single numerical value, similar to MOS, that relates to the degree of artifacts produced by a VC system $s$. To this end, we replace the abstract human observer $h(\mathcal{Y}, \mathcal{O}_i)$ by a \emph{machine} observer, $m(\mathcal{Y},\mtx{\theta})$, represented by some model parameters $\vec{\theta}$. There are several differences between $h$ and $m$. First, unlike $h$, evaluating $m$ is \emph{deterministic} and \emph{time-invariant} --- in other words, it yields the same output when repeated on sample $\mathcal{X}$, and is not dependent on the time when invoked. Second, unlike $h$ where one usually constrains the outputs to be quantized to a small set of ordinal variables, we allow the domain of $m$ to be the entire real line $\mathbb{R}$; the scale of the output value is arbitrary, but similar to $h$, higher numerical values in relative terms indicate higher speech quality as judged by the observer $\vec{\theta}$.

To flesh out the above vague idea, in this work $m(\cdot,\vec{\theta})$ takes the form of a \emph{likelihood ratio detector}. Likelihood ratios arise naturally from the Bayes theorem and serve as the starting point for making statistically optimal decisions. For a given input utterance $\mathcal{Y} \in \mtx{Y}$, we compute a \emph{log-likelihood ratio} (LLR) score,
	\begin{equation}\label{eq:detector-llr}
    	\ell(\mathcal{Y}|\vec{\theta}) = \log \frac{p(\mathcal{Y}|H_0)}{p(\mathcal{Y}|H_1)} = \log \frac{p(\mathcal{Y}|\vec{\theta}_\text{nat})}{p(\mathcal{Y}|\vec{\theta}_\text{artif})},
    \end{equation}
where $\vec{\theta}=(\vec{\theta}_\text{nat}, \vec{\theta}_\text{artif})$. The null hypothesis $H_0$, modeled through $\vec{\theta}_\text{nat}$, is that $\mathcal{Y}$ represents natural human speech without artifacts. The alternative hypothesis $H_1$, modeled using $\vec{\theta}_\text{artif}$, states that $\mathcal{Y}$ originates from artificial speech generation (such as voice conversion or speech synthesis). Higher numerical values are therefore associated with speech that appears more `human-like', lacking vocoding artifacts or other problems that VC systems tend to generate.

To train $\vec{\theta}_\text{nat}$, we gather a large collection of natural human utterances and train the model from the pooled data; similarly, to train $\vec{\theta}_\text{artif}$, we gather a representative collection of artificial speech samples (such as samples from several state-of-the-art VC systems, \emph{prior} to evaluating a new VC system). In this work, we use Gaussian mixture model (GMM) to model each hypothesis, trained through expectation-maximization (EM) algorithm. Though more advanced spoofing countermeasure backends are available, GMMs produced good results for the ASVspoof'15 challenge consisting also of high-quality clean samples as here, with the benefit of simplicity.

\subsection{Error rate of the CM as an objective quality measure}

The log-likelihood ratio in \eqref{eq:detector-llr} outputs a number for a single utterance $\mathcal{Y}$. Now, how do we obtain a summary value for all the samples of a given VC system $s$? While it might be appealing to just average $\ell(\mathcal{Y}_n|\vec{\theta})$ over the samples $\mathcal{Y}_n \in \mtx{Y}^s$, similar to MOS, this is not recommendable. Unlike the opinion score, $\ell$ is unbounded and is therefore more difficult to interpret. Further, the model $\vec{\theta}$ is imperfect version of reality and cannot possibly produce meaningful LLR scores for all human speech and all spoofing attacks. As a result, the scale of $\ell$ is arbitrary and dependent on the various modeling choices (including that of feature extraction). The LLRs across databases, VC systems and converted utterances are not in a commensurable scale. In the nomenclature of ASV literature, we might say $\ell$ is not \emph{well-calibrated} \cite{Leeuwen2013-calibration}.

Hence, it is better to adopt a summary measure that is more intuitive and comparable across different evaluation environments. Our proposal is the \emph{error rate} of the detector, as a measure of its ability to differentiate authentic human utterances from those generated by voice conversion. Our philosophy is as follows. If the samples from a VC system $S_1$ manage to fool a given artificial speech detector more often than samples from another competitive VC system $S_2$, we can say samples of $S_1$ appear more human-like in the eyes\footnote{Or ears?} of the artificial speech detector. Utterances having less processing artifacts are more difficult to discriminative from human speech, giving rise to higher error rate of the detector.

Even if our inspiration for such a proposal originates from the work in ASV anti-spoofing, the viewpoint is now switched from \emph{defender} to \emph{attacker}. In the ASV anti-spoofing, one keeps improving spoofing countermeasures so that they are more accurate in detecting advanced speech synthesis and voice conversion attacks (the lower the error rate, the better). But now we consider the anti-spoofing system to be fixed with the goal of improving the performance of voice conversion systems --- the higher the error rate, the better. 

\begin{table*}[!t]
\caption{Considered datasets for the construction of spoofing countermeasures for objective quality assessment of the VCC'18 samples. The datasets differ in the number of speakers and diversity of spoofing attacks. The ASVspoof'15 data represented state-of-the-art of SS and VC in 2014-2015, while VCC'16 contains more modern attacks. \textbf{ASVspoof:} \emph{Automatic Speaker Verification Spoofing and Countermeasures Challenge}, \textbf{VCC:} \emph{Voice Conversion Challenge}, \textbf{SS:} \emph{speech synthesis}, \textbf{VC:} \emph{voice conversion}, \textbf{STRAIGHT}: \emph{Speech Transformation and Representation using Adaptive Interpolation of
weiGHTed spectrum}, and \textbf{LPC}: \emph{linear predictive coding}.
}
\label{tab:cqcc-datasets}
\begin{center}
\begin{tabular}{lccccc}
								& \textbf{ASVspoof'15 train} 	& \textbf{ASVspoof'15 dev} 		& \textbf{ASVspoof'15 eval} 		& \textbf{VCC'16}	& \textbf{VCC'18 base}\\
\hline\hline
Types of attacks 				& SS and VC 					& SS and VC 					& SS and VC 						& VC				& VC \\
Waveform generation 			& STRAIGHT,  					& STRAIGHT 			& STRAIGHT, diphone					& STRAIGHT, LPC,	& Waveform \\
								& MLSA							&  								& concatenation 					& Ahocoder			& filtering\\
\# Spoofing attacks 			& 5 							& 10 							& 10 								& 18				& 1 (VCC'18 basel.) \\
\# Spoof files 					& 12,625 						& 49,875						& 184,000 							& 24,300			& 2,240 \\
\# Speakers (\male + \female) 	& 25 (10 + 15)					& 35 (15 + 20) 					& 46 (20 + 26) 						& 10 (5+5)			& 8 (4 + 4)\\
\# Human files 					& 3,750 						& 3,497 						& 9,404 							& 535				& 464 \\
\hline
\end{tabular}
\end{center}
\end{table*}

As for the actual error rate measurement, we adopt the standard metric of \emph{equal error rate} (EER) used extensively in ASV, anti-spoofing and biometrics research. The output scores (estimated LLRs) produce two different types of errors, \emph{false alarms} and \emph{misses}, that are traded off with respect to each other. Here, false alarm (false acceptance) rate (FAR) is the proportion of artificial speech samples that the detector falsely accepted as \emph{bona fide} (human) samples. Miss (or false rejection) rate (MR), in turn, is the proportion of falsely rejected bona fide samples. FAR and MR are, respectively, decreasing and increasing functions of a detection threshold. The EER, then, is the unique error rate corresponding to the threshold at which FAR and MR equal each other. As the detection task involves only two classes, the chance level is EER of 50\%. This would be our `ideal' value for a successful artifact-free VC system\footnote{Technically speaking it is possible to do \emph{worse} than the coin-flipping rate, for instance by swapping the two model likelihoods in \eqref{eq:detector-llr}. EERs much larger than 50 \% suggest usually an implementation bugs of the detector, and are not interesting from the perspective of evaluation.}. One might therefore optionally report a scaled version $\frac{1}{10}\text{EER \%}$, to give rise to a continuous version of a 5-point `opinion score', judged by the machine observer.



\subsection{Choice of the countermeasure model \eqref{eq:detector-llr}}

To implement the artifact LLR detector of \eqref{eq:detector-llr}, we represent speech utterances using a sequence of short-term spectral features, $\mathcal{Y}=\{\vec{y}_1,\dots,\vec{y}_N\}$ with $\vec{y}_n \in \mathbb{R}^D$, $D$ being the feature dimensionality. At the training stage, we use the pooled feature vectors from each class to independently train two Gaussian mixture models (GMMs), $\vec{\theta}_\text{nat}$ and $\vec{\theta}_\text{artif}$, using the standard \emph{expectation-maximization} (EM) algorithm. We use diagonal covariance matrices and consider the number of Gaussian components, $C$, as a tuning parameter. It can be used to adjust the balance between over- and under-fitting. 

Choice of the short-term feature representation is critically important in the context of spoofing countermeasures \cite{Sahid2015-features}. Findings from the first ASVspoof 2015 challenge \cite{Wu2017-asvspoof} highlighted the importance of spectral and temporal \emph{details} for the task of discriminating real and spoofed speech. In specific, conventional MFCCs, a \emph{low-resolution}, \emph{low-frequency focusing} feature set that has enjoyed its \emph{de facto} audio representation status for almost 40 years, is a suboptimal choice. Discriminating human speech from synthetic or converted speech and identifying the artifacts seems to require more detailed time-frequency representation. The winning system of the ASVspoof'15 challenge \cite{Patel15-combining} used MFCCs with a combination of cochlear filter cepstral coefficients and instantaneous frequency. Later, \cite{Todisco2016-cqcc} introduced a single feature set, \emph{constant-Q cepstral coefficients} (CQCCs), based on the constant-Q transform \cite{Brown91}. 
It lead to the lowest reported EERs on the ASVspoof'15 corpus at the time. Substantial follow-up work (\emph{e.g.} \cite{Sriskandaraja-scattering-2017}) has improved feature extractors even further. In this work we use CQCCs due to their high reported detection accuracy, simplicity, and widespread adoption by the research community. We use an open-source CQCC implementation provided to the second ASVspoof challenge participants\footnote{\url{http://www.asvspoof.org/data2017/baseline_CM.zip}}, and similarly another public toolkit to train the GMMs \cite{msrtoolkit2013}.

\subsection{Data-related considerations to enable fair evaluation}

Besides specification of the front-end features, another key consideration is the choice of training and development data of the countermeasure. Despite high accuracy of the spoofing countermeasure front-ends listed above, they are notoriously sensitive to training-test mismatch (cross-corpus performance) \cite{Korshunov2017-xdb-score-fusion}, additive noise \cite{hanilci16-noisy} and channel/bandwidth mismatch \cite{Delgado17-bandwidth}. In short, countermeasures are easy to overfit for specific data leading to potentially arbitrarily bad results for a different test data. This makes the selection of both data, and optimization process of the countermeasure parameters important.

As challenge organizers, our responsibility is to avoid favoring any specific VC system but provide as unbiased assessment of the systems as we possibly can. To this end, we aim to optimize our countermeasure with the following requirements in mind. 
\begin{enumerate}
	\item \textbf{Stability across datasets.} The countermeasure should show stable enough results when executed on different corpora so that one can trust the result to be less dependent on the specifics of VCC'18 data.
    \item \textbf{Detection of state-of-the-art voice conversion.} We should aim at detecting the current state-of-the-art, or otherwise previously known, VC attacks, with sufficient accuracy. 
    \item \textbf{No tweaking using participant submissions.} We should \emph{not} optimize our proposed measure with feedback of the VCC'18 evaluation entries. That is, we should not look at the error rates and use the submitted samples to enrich training sets. Instead, we should fix the data and parameters to the best-known values using \emph{data that precedes VCC'18 participant entries}.
\end{enumerate}

The last requirement might be less obvious to the reader from the voice conversion field. It reflects the viewpoint of the spoofing countermeasure as a security gate in real-world deployment: one does not know the attacks (here, voice conversion samples) in advance but has to use her best knowledge to prepare the countermeasure using attacks available beforehand. In the standard automatic speaker verification (ASV) evaluation benchmarks conducted by National Institute of Standards and Technology (NIST), evaluation participants are similarly expected to process the new evaluation samples completely blindly --- they are not allowed to interact with the evaluation samples in any manner (such as by listening to them, or using them to do modeling decisions), for the same reason. We apply the same principle in our role as a challenge evaluator, to give all the submitted VCC'18 systems an equal opportunity to break down our countermeasure.

\section{Experimental data}

\subsection{The voice conversion challenge 2018}

The 2018 Voice Conversion Challenge (VCC'18) is a follow-up to the VCC series kicked off in 2016 \cite{Toda2016-VCC-challenge}. It features the task of speaker identity conversion. A detailed description of the challenge data, rules, analysis of the submitted systems and extensive perceptual results are provided in another paper \cite{Lorenzo2018-VCC18}, with only the key facts repeated here for completeness. 

The VCC'18 challenge is designed to promote development of both \emph{parallel} and \emph{nonparallel} VC methods; the parallel (\textbf{Hub}) task contains source-target training utterances with matched speech content while in the nonparallel (\textbf{Spoke}) task, the contents differ. Both tasks contain the same target speaker data but the source speakers are different. The Hub task formed the core (required) task for the registrants, whereas participation to the Spoke task was optional. The participants were provided with training and development data, and were asked to submit their converted audio files for previously unseen source utterances.


Both the VCC'16 and the VCC'18 data are based on the DAPS (Data And Production Speech) dataset \cite{DAPS} including native US English speakers recording in a professional setting. The source and the target speakers across the two challenges are all disjoint. The number of test sentences is 35 and the participants were asked to submit the converted voices for a total of 16 source-target speaker pairs in both Spoke and Hub tasks. The results were evaluated subjectively using crowd-sourcing. A total of 267 unique crowdworkers evaluated perceptually both naturalness and speaker similarity. The former ranged from 1 (completely unnatural) to 5 (completely natural) while a 4-point scale was used for the latter (``Same, absolutely sure'', ``Same, not sure'', ``Different, not sure'', ``Different, absolutely sure''. The trials consisted of comparisons of VC samples with either the source speaker or the target speaker.

\subsection{Data for countermeasure development}

With the above considerations in mind, we involve data from several audio collection that contain both synthetic speech and converted voice samples. The datasets are summarized in Table \ref{tab:cqcc-datasets}. As for the \textbf{ASVspoof 2015} collection (train, dev, eval), documented in detail in \cite{Wu2017-asvspoof}, we follow the protocol files provided with the corpus\footnote{\url{https://datashare.is.ed.ac.uk/handle/10283/853}}. The \textbf{VCC'16} data, in turn, consists of the samples of the first edition of the VCC series \cite{Toda2016-VCC-challenge}. In specific, it contains the participant submissions from 17 different systems plus 1 baseline system, along with samples from 10 speakers (three source males, two source males, two target females, three target males). 

The last data, denoted \textbf{VCC'18 base}, contains \emph{only} the VCC'18 baseline as its only attack (in specific, samples of the VCC'18 baseline system). The human trials contain speech from 8 speakers in the VCC'18 challenge. The four source speakers used in the HUB task of VCC'18 have an overlap with VCC'16 data and are excluded from the trials. Again no submitted VCC'18 system is used for training the countermeasures. 


\begin{table}[!t]
\caption{Equal error rate (EER, \%) for intra-corpus (ASVspoof'15) and cross-corpus artifact detection experiments, the \emph{lower} the better. Number of Gaussians 32, 29 CQCCs and energy coefficient with deltas and double deltas. The EERs are averages of attack-specific EERs.}
\label{tab:cqcc-datasets}
\begin{center}
\begin{tabular}{lll|ll}
&  						&					& CQCC 			& CQCC\\
& Train						& Test 							& (raw) 	& (CMVN)\\
\hline\hline
(i) & TRAIN'15		& DEV'15 	& 0.38 						& 1.99\\
(ii) & TRAIN'15 		& EVAL'15 	& 1.83 						& 2.21\\
\hline
(iii) & TRAIN'15 		& VCC'16 	& 35.04 					& 34.26\\
(iv) & ALL'15 			& VCC'16 	& 34.86						& 31.52\\
(v) & VCC'16 			& DEV'15 	& 26.18 					& 33.34\\
(vi) & VCC'16 			& EVAL'15 	& 21.48 					& 34.75\\
\hline
\end{tabular}
\end{center}
\end{table}

\section{Results}

\begin{table}[!t]
\caption{Equal error rate (EER, \%) for CQCC optimization, the \emph{lower} the better. Number of Gaussians 32, training data VCC'16. Lines (i) to (vii) are based on 29 CQCCs without the energy coefficient while (viii) contains the zeroth coefficient. The EERs are averages of attack-specific EERs.}
\label{tab:delta-opt}
\begin{footnotesize}
\begin{center}
\begin{tabular}{ll|ll|ll|}
&								& \multicolumn{2}{c|}{\textbf{DEV'15}}		& \multicolumn{2}{c|}{\textbf{VCC'18 base}}\\
&								& CQCC 				& CQCC					& CQCC			 & CQCC\\
& Front-end						& (raw) 			& (CMVN)				& (raw) 		 & (CMVN)\\
\hline\hline
(i) & stat						& 46.14				& 32.33					& 28.50			 & 32.67\\
(ii) & $\Delta$ 				& 12.11 			& 31.20					& 32.45 		 & 33.15\\
(iii) & $\Delta^2$ 				& 13.58 			& 29.39 				& 30.96 		 & 31.74\\
(iv) & $\Delta,\Delta^2$ 		& \textbf{7.73} 	& \textbf{18.93}		& 30.30 		 & \textbf{26.75}\\
(v) & stat, $\Delta$ 			& 39.73 			& 36.86					& 27.54			 & 35.07\\
(vi) & stat, $\Delta^2$ 		& 34.46 			& 31.65	 				& 24.36 		 & 31.14\\
(vii) & stat, $\Delta,\Delta^2$ & 32.09 			& 34.28					& 25.01 		 & 31.04\\
(viii) &z, stat, $\Delta,\Delta^2$	& 26.18 		& 33.34 				& \textbf{24.15} & 27.91\\
\hline
\end{tabular}
\end{center}
\end{footnotesize}
\end{table}

%

\begin{figure}[t!]
    \centering
\includegraphics[scale=0.36]{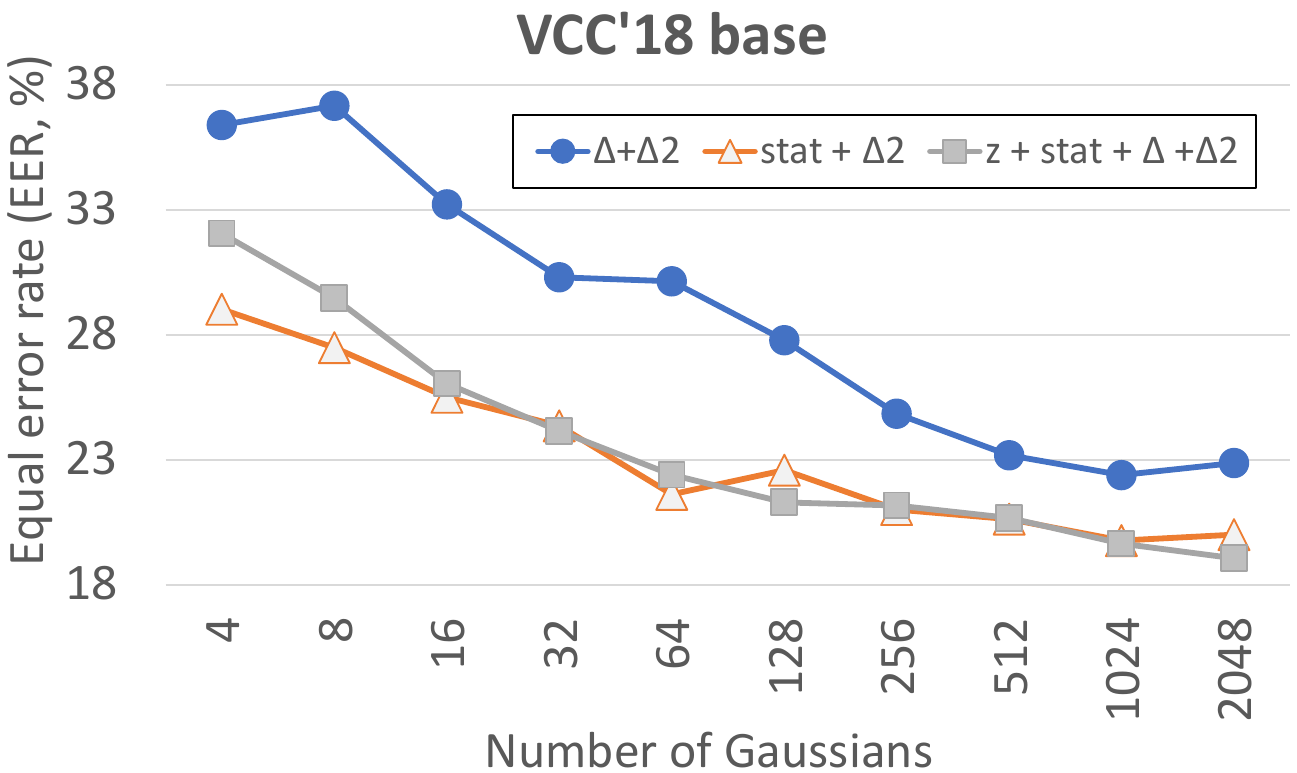}
    \caption{The results on VCC'18 base data with varied complexity of the GMM backend. Training data is VCC'16.}
    \label{fig:gaussian-count}
\end{figure}

\subsection{CQCC-GMM countermeasure optimization}

Given the sampling rate mismatches across the prior corpora (16 kHz for ASVspoof'15 and VCC'16) and the new VCC'18 data (22.05 kHz), we downsample the latter to 16 kHz. In our first experiment, we study the performance of the CQCC-GMM detector for different selections of training and test data, with the primary goal being selection of our main training data. For this first experiment, we fix the CQCC configuration to the default setting of the ASVspoof'17 challenge baseline (29 base CQCCs plus the zeroth (energy) coefficient, with deltas and double deltas, giving  90-dimensional features. We study the CQCC configurations both without any feature normalization, as well as with cepstral mean and variance normalization (CMVN); it was included since it might be potentially helpful in suppressing convolutive mismatch across datasets. Convolutive bias could originate from speaker, recording media or vocoder differences, and might be reducible through feature normalization techniques. In our implementation, we use utterance-level CMVN to obtain zero mean, unit variance features per file. We do not apply speech activity detection. The number of Gaussian components is set to $32$.

\begin{table*}[!t]
  \centering
  \caption{Equal error rates (EER \%) of the CQCC-GMM spoofing countermeasure of the VCC'18 entries on the Hub task. Here ``B01" denotes the VCC'18 baseline system. The two countermeasures considered use deltas and double deltas ($\Delta+\Delta^2$) of 29 CQCCs, and 29 CQCCs plus zeroth coefficient along with deltas and double deltas (All feat.). Training data VCC'16, number of Gaussians 2048. The \emph{higher} the EER, the better the VC system in terms of quality (less processing artifacts). ?: information unavailable to the authors.}
    \label{tab:vcc18-eer}
    \begin{tabular}{c c c c | c c c c}
        \hline
         				& Waveform 		& $\Delta +\Delta^2$ 	& All feat. 	&   			& Waveform 		& $\Delta +\Delta^2$ 	& All feat. \\
        \textbf{Sys.}	& generation 	& 						& 				& \textbf{Sys.} & generation \\
        \hline\hline
        B01 			&  Waveform filtering  & 29.77 & 25.55 		&  N09 & Ahocoder & 1.24 & 13.22        \\
        D01 			&  World direct wave mod. & 8.43  & 18.75  	&  N10 & Wavenet & 4.64 & 15.63       \\
        D02 			&  World  & 43.94 & 48.95 					&  N11 & STRAIGHT & 0.49 & 1.28      \\
        D03 			&  STRAIGHT & 1.58  & 16.07  				&  N12 & Waveform filtering & 40.98 & 23.48     \\
        D04 			&  World & 15.74  & 15.03  					&  N13 & World & 10.69 & 17.42      \\
        D05 			&  World & 4.03  & 16.63  					&  N14 & Waveform filtering  & 41.58 & 25.95         \\
        N03 			&  ? & 5.80  & 16.47  						&  N15 & World & 7.24 & 17.18          \\
        N04 			&  World & 3.74  & 16.26  					&  N16 & Ahocoder & 1.18 & 11.17     \\
        N05 			&  SuperVP & 41.42 & 32.23 					&  N17 & Wavenet & 15.19 & 15.48     \\
        N06 			&  World & 2.72  & 14.55 					&  N18 & Griffin-Lim & 32.35 & 19.43     \\
        N07 			&  World & 4.12  & 16.75  					&  N19 & World & 11.07 & 21.97         \\
        N08 			&  Waveform filtering & 37.20 & 23.84 		&  N20 & World & 3.54 & 15.23         \\
        \hline
    \end{tabular}
\end{table*}

The results for the ASVspoof'15 and VCC'16 data are shown in Table \ref{tab:cqcc-datasets}. The first two rows correspond to the standard protocols of the ASVspoof'15 corpus and reflect intra-corpus performance. The results are in line with the published literature. First, the error rates are remarkably low, demonstrating the potential of the CQCC-GMM countermeasure. Second, the error rate of the evaluation part is higher, due to presence of one \emph{unknown} attack (S10). CMVN systematically degrades performance. 

The last four rows of Table \ref{tab:cqcc-datasets} show the cross-corpus performance. As expected, the error rates are now far higher. The training set ALL'15 indicated in line (iv) was obtained by pooling the train, dev and eval files of ASVspoof'15 into a large training set. Comparing experiments (iii) and (iv), the larger training gives no substantial boost for the unnormalized features (relative decrease of 0.5\% in EER) though with some improvement (8\% relative decrease) for the normalized features. CMVN is helpful in (iv) only. Comparing experiments (v) and (vi) to (iii) and (iv) of the unnormalized features, the results are not symmetric regarding the roles of training and test corpus. Even if the VCC'16 data is much smaller than ASVspoof'15 in terms of speaker and file count, the voice conversion samples (attacks) are perhaps more diverse acoustically. 

Based on the results of Table \ref{tab:cqcc-datasets}, for the remainder of the experiments we fix VCC'16 as our training data. The next experiment concerns the impact delta features that have been noted accurate in detecting vocoded speech.  Table \ref{tab:delta-opt} shows the results for the ASVspoof'15 dev (the more \emph{difficult} one from dev and eval), and VCC'18 base. The results are shown again for raw and CMVN-processed features. 

For the ASVspoof'15 dev trials, the outstanding front-end consists of just deltas and double deltas without feature normalization. The results highlight usefulness of the dynamic features, and \emph{un}usefulness of the static coefficients --- the first line consisting of static coefficients only gives performance close to the chance level of 50\% EER. Comparing experiment (ii) to (v), experiment (iii) to (vi) and experiment (iv) to line (vii), inclusion of the static coefficients systematically degrades performance. It is also noteworthy that while the plain delta features are degraded by CMVN, CMVN boosts the performance of the static coefficients in cases (i), (v) and (vi). This might be partly explained noting that both CMVN and deltas are helping in reducing convolutive mismatch. 

For the VCC'18 base data, CMVN systematically degrades performance; the only exception is the configuration consisting of deltas and double deltas only. The globally best setups for both trial sets are obtained without CMVN. Comparing the best configurations from each trial set, 7.73\% for the ASVspoof'15 dev and 24.15\% for VCC'18 base, the latter appears harder. This suggests that the state-of-the-art voice conversion attack (VCC'18 baseline) is challenging for the CQCC-GMM countermeasure. 


In our last parameter tweaking experiment, we fine-tune the number of Gaussians (that was 32 until now for computational reasons) using just the VCC'18 base data (training with VCC'16). Based on the results of Table \ref{tab:delta-opt}, we select three representative feature set-ups, the one that produced good results on the ASVspoof'15 dev data; and two good set-ups (stat + $\Delta^2$ and z + stat + $\Delta$ + $\Delta^2$) for the VCC'18 base data. The results displayed in Fig. \ref{fig:gaussian-count} indicate improved performance with larger number of Gaussians as expected. The performance might slightly be improved by increasing the number of Gaussians further; due to resource reasons, we stopped at 2048. Concerning the front-end set-up, the full configuration containing static, delta, double delta and the zeroth coefficients yields the best results. 

\subsection{Results for the VCC'18 samples}

We now fix our countermeasure parameters to compare the VCC'18 submissions. From Table \ref{tab:delta-opt}, we see that CMVN helps only in 3 (out of 16) cases so we decide to not include it. Based on Table \ref{tab:delta-opt}, the $\Delta+\Delta^2$ is a reasonable choice of features: it yields a clearly outstanding result on the difficult cross-corpus experiment with ASVspoof'15 dev and, with optimized number of Gaussians (Fig. \ref{fig:gaussian-count}), works reasonable well also for the VCC'18 base data.  Additionally, we include the full feature set-up consisting of base coefficients (including energy) with deltas and double-deltas, as this yielded the best overall results in the detection of the baseline samples. Based on Fig. \ref{fig:gaussian-count}, we fix the number of Gaussians to $2048$.

The results are shown in Table \ref{tab:vcc18-eer} for both of the selected feature setups on the VCC'18 Hub task. For ease of interpretation, we show the waveform generation method of each submission entry. For a case of the $\Delta+\Delta^2$ configuration, waveform filtering, SuperVP and Griffin-Lim based waveform generation methods were judged as VC methods that have relatively less artifacts compared to STRAIGHT, World, and Ahocoder vocoders. This is reasonable since they were proposed for improving issues of minimum phase vocoders. One suprising result to everyone may be that althought N10 was evaluated as the best VC by human listeners (about 4.1 MOS score), our methods detected its artifacts easily and its EER is low as 4.6 \%. The N10 does not use any deterministic vocoders as above, but, uses $\mu$-law quantized waveforms instead \cite{oord2016-wavenet}. The $\mu$-law quantization may cause obvious artifacts (although they are non audible to human and hence they were well evaluated by human listeners). 

One interesting exception to our expectations is system D02. For the $\Delta+\Delta^2$ configuration, the VC system D02 has achieved highest EER although they have used the known vocoder. According to the listening test \cite{Lorenzo2018-VCC18}, the D02 samples sound very similar to source speakers and perfectly dissimilar to target speaker. This suggests that D02 used little modification of source speaker's waveforms and hence has likely less processing artifacts.  

We can also see that the EERs are sensitive to the choice of acoustic features used. Using the full feature configuration, which was optimized to detect the VCC'18 baseline samples, the waveform filtering, SuperVP and Griffin-Lim based waveform generation methods still have high EERs, but there are also classic systems, such as N19, that obtain relatively high EERs despite being known technology. We suspect that the countermeasures are overfit and incapable of generalizing beyond the training data. It would be important for us to develop stable and robust models. Having said that, majority of the VC systems do not achieve EERs close to the chance level (50\%) regardless of the selected features, indicating that all the VC methods produced artifacts more or less and having plenty of margin for improving the quality of the samples.

\begin{figure*}[t!]
    \centering
  \includegraphics[scale=0.32]{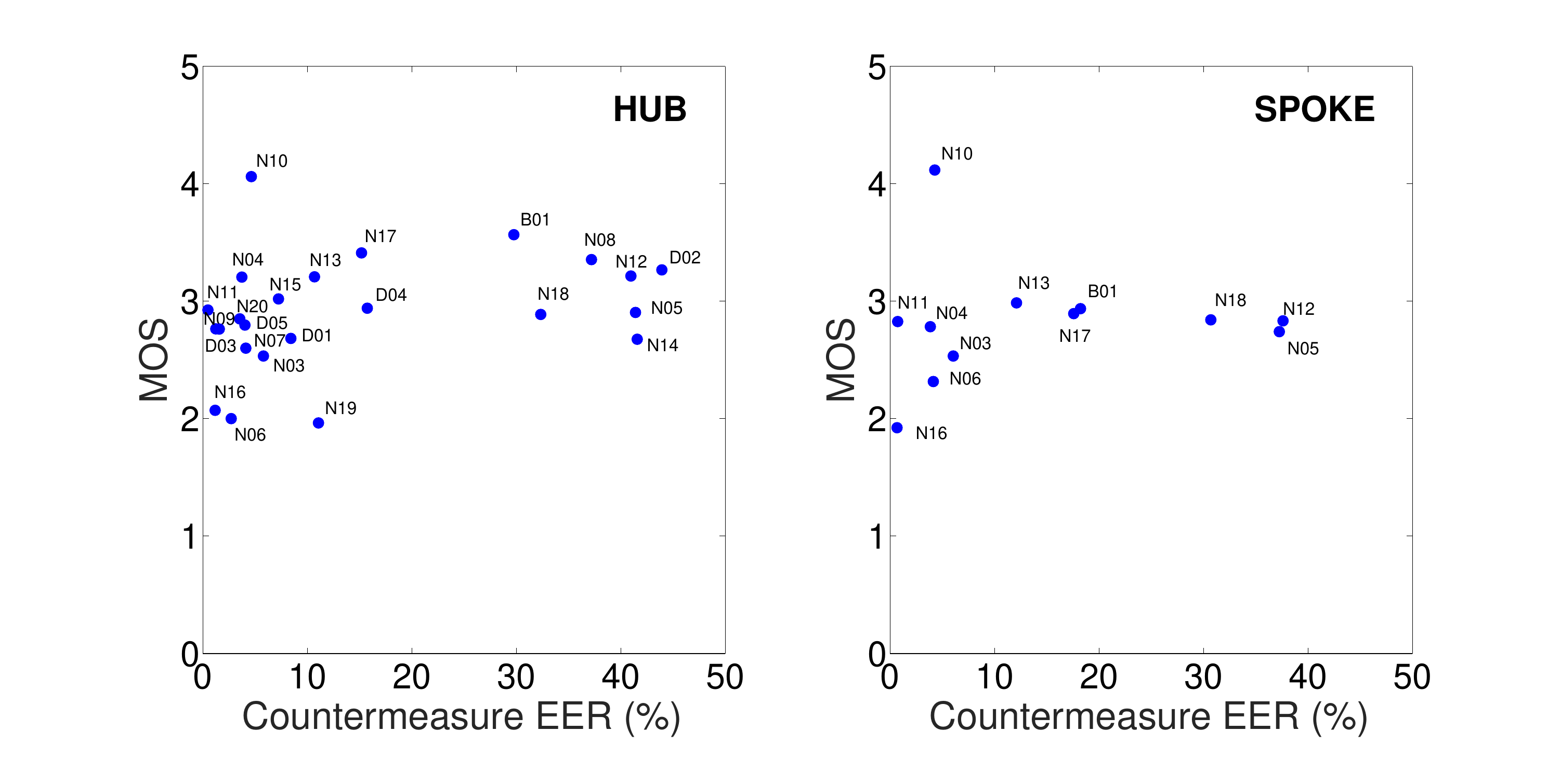}
    \caption{Scatter plot of objective vs. subjective quality of the VCC'18 challenge entries for the HUB and SPOKE tasks. The vertical axis represents mean opinion score (MOS) of subjective quality ratings of the samples of each challenge entry, while the horizontal axis represents equal error rate (EER, \%) of spoofing countermeasure optimized to detect human and artificial speech. Therefore, the \emph{higher} the EER value, the more confused the countermeasure is in telling apart the converted samples from authentic human speech, implying higher quality of the samples. The `ideal' values would be MOS = 5.0 and EER = 50 \%.}
    \label{fig:cqcc-vs-MOS}
\end{figure*}

Finally, Fig. \ref{fig:cqcc-vs-MOS} displays a scatter plot of the EER of the $\Delta+\Delta^2$ countermeasure against the mean opinion score (MOS) from the perceptual experiments detailed in \cite{Lorenzo2018-VCC18}. We do not observe strong association between the two but they are complementary to each other. This is not  entirely surprising remembering that our objective measure focuses on audible and non-audible artifacts, whereas listeners evaluated audible naturalness subjectively. As expected, human perception and machine perception are different.  




\section{Conclusion}

We have proposed the use of spoofing countermeasure for objective artifact assessment of converted voices as a supplement to the VCC'18 challenge results. Our approach is reference-free and text-independent in the sense that it does not require access neither to the original source speaker waveform nor any text transcripts. It assigns a single EER number to a batch of converted speech utterances from a single VC system, and can therefore be used in comparing different VC systems in terms of their artifacts. 

Although the tested countermeasure utilizing CQCC front-end and GMM backend was found to be sensitive to the choices of model parameters and acoustic features, our results indicate clear potential of spoofing countermeasure scores as a convenient and complementary tool to automatically assess the amount of audible and non-audible speech artifacts.

The obtained results are in a reasonable agreement with types of waveform generation methods used for VC systems. Our results indicate that the waveform filtering, SuperVP and Griffin-Lim methods have relatively less artifacts. Since they are not included in the current ASVspoof datasets, we claim that it is also important to create new spoofing materials based on the methods and to train more robust anti-spoofing countermeasures in practice. 

We also revealed that perceptually convincing VC samples based on Wavenet \cite{oord2016-wavenet} in the VCC'18 have detectable artifacts. This implies that the current best VC samples may fool human ears but not necessarily the CM systems. There is no system that would perfectly fool both the humans and CM systems yet.

While our study serves as a proof-of-concept, we foresee several possible future directions. Firstly, it would be interesting to compare the performance to standard objective artifact measures used in assessing speech codecs and speech enhancement methods, and to other spoofing countermeasure front-ends besides CQCCs. Some alternative features may provide a stronger correlation to the MOS scores. Second, given the obvious issue related to selection and enriching of the training data with the latest VC techniques it might be relevant to consider one-class approaches that require human training speech only. Even if such approaches have had only moderate success in anti-spoofing \cite{Alegre2013-one-class} it  would be interesting to revisit them in the context of artifact assessment.

\section{Acknowledgements}
We are grateful to iFlytek Ltd.\ for sponsoring the evaluation of the VCC 2018. This work was partially supported by MEXT KAKENHI Grant Numbers (15H01686, 16H06302, 17H04687, 17H06101), and Academy of Finland (proj. no. 309629).



\bibliographystyle{IEEEbib}


\end{document}